\documentclass[mathleft
]{an}
\usepackage{epsfig}
\usepackage{epstopdf}
\usepackage{graphicx}
\usepackage{times}
\usepackage{color}

\usepackage{natbib}
\begin{document}
\sloppy
\Pagespan{789}{}
\Yearpublication{2010}%
\Yearsubmission{2010}%
\Month{11}%
\Volume{999}%
\Issue{88}%

\title{Asteroseismic study of $\gamma$\,Doradus members of the cluster NGC 2506}

\author{A. Grigahc{\`e}ne\inst{1}\fnmsep\thanks{Corresponding author:
  \email{ahmed.grigahcene@astro.up.pt}\newline}, A. Moya\inst{2}, J.-C. Su{\'a}rez\inst{3} \and S. Mart{\'{\i}}n-Ruiz\inst{3}
}


\titlerunning{Asteroseismic study of $\gamma$\,Dor NGC 2506 V$_{11}$}
\authorrunning{Grigahc{\`e}ne et al.}
\institute{
Centro de Astrof\'{\i}sica, Faculdade de Ci\^encias, Universidade do Porto, Rua das Estrelas, 4150-762 Porto, Portugal
\and 
Laboratorio de Astrof\'isica Estelar y Exoplanetas, LAEX-CAB (INTA-CSIC), PO BOX 78, 28691 Villanueva de la Ca\~nada, Madrid, Spain
\and 
Instituto de Astrof\'{i}sica de Andaluc\'{i}a (CSIC), CP3004, Granada, Spain
}
\received{2010}
\accepted{2010}
\publonline{later}


\abstract{The present work performs an asteroseismic study of a
          $\gamma$\,Doradus star member of the open cluster NGC~2506.
          This works conjugates the observational information of the
          star and the strong constraints imposed by the cluster
          memebership, with the latest developed theoretical
          asteroseismic tools for the modelling of this type of objects,
          in particular the Frequency Ratio Method (FRM)
          and Time-Dependent Convection (TDC), to study the $\gamma$\,Dor
          stars found in the NGC 2506 stellar cluster. The
          use of both techniques gives us the opportunity of constructing a
          self-consistent procedure, allowing the mode identification and
          improving the modelling of the $\gamma$\,Dor members of the cluster. 
          In this work we present the result of the analysis of the first
          target of our project, showing the advantadge of modelling
           cluster member's $\gamma$\,Dor stars. This is the first step in a more ambitious project. In the near future, we plan to expand the study to other variable members of the NGC 2506 cluster. This will improve our knowledge of $\gamma$\,Dor pulsatorsand probe the self-consistency of our procedures.}

\maketitle

\section{Introduction}
The $\gamma$\,Doradus (Dor) stars are suitable objects for astroseismic study. They are high-order gravity ($g$)~mode
  pulsators, which permits the probing of their stellar coreand the
  use of the asymptotic approximation, for which analytical solutions
of the oscillation equations exist (\citealt{Smey07}). The Frequency
Ratio Method (FRM; \citealt{Moya05, Suarez05}), based on this property,
is particularly adapted for obtaining asteroseismic information on 
$\gamma$\,Dor pulsating stars that show, at least, three pulsation frequencies. 
The method provides an estimate of the identification of the radial order
$n$ and the spherical degree $\ell$ of observed frequenciesand
of the integral of the buoyancy frequency (Brunt-V\"ais\"al\"a)
weighted over the stellar radius along the radiative zone ($I_{\rm{obs}}$).

The excitation mechanism of $\gamma$\,Dor
stars was an outstanding problem until the works of \cite{Guzik00} and
\cite{Dupt05} who used respectively, frozen and Time-Dependent Convection (TDC) theories
showed that the position of the base of the
convective envelope is the key for driving $\gamma$\,Dor $g$~modes. 
Moreover, TDC is formulated in the framework of mixing length theory and therefore the
$\gamma$\,Dor instability strip is predicted to be sensitive to the mixing-length
parameter $\alpha_{\rm MLT}$.

The asteroseismic modelling of $\gamma$\,Dor, even armed with these techniques, still needs the global parameters of the observed stars. One of the ways to obtain a significant
enhancement is to use the strong constraints imposed by the cluster membership for example, the age,
metallicity and distance (\citealt{2002A&A...390..523S, 2007MNRAS.379..201S,2009A&A...507..901C}).

At the beginning of the recognition of the $\gamma$\,Dor stars as a new class of
variable stars, a great effort has been devoted to their detection in clusters. At that time the main aim of the
  community was to check whether this type of pulsations was linked or not to age and obviously,
stellar clusters were the most suitable targets for this purpose (\citealt{2002ASPC..259..152M}).

One of the richest cluster in $\gamma$\,Dor members is the old open cluster NGC 2506, for which photometric observations 
 provide a considerable number of variable stars (\citealt{2007A&A...465..965A}). Among them, 15 have been identified as $\gamma$\,Dor
stars candidates (labeled as V$_{11}$ -- V$_{25}$ in that paper).

This work attempts to investigate the impact of using a procedure
which includes FRM and TDC on the modelling of
$\gamma$\,Dor, stars taking advantage of their membership in clusters. In particular, we here provide preliminary
results for the $\gamma$\,Dor star V$_{11}$, belonging to the open
cluster NGC 2506.

\section{Observational Data}

\begin{figure}[t]
\centering
\includegraphics[width=82mm,height=90mm]{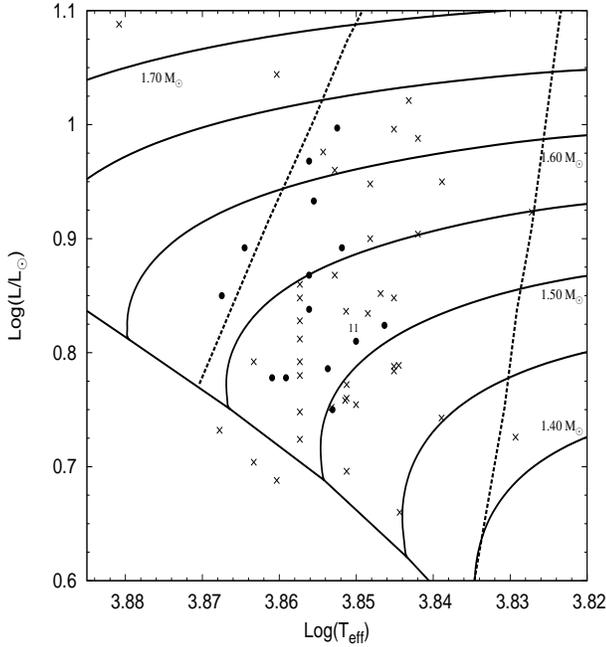}
\caption{HR Diagram position of $\gamma$\,Dor variables. Crosses represent the stars parameters given by \cite{Hen07}. The NGC 2506 $\gamma$\,Dor members (\citealt{2007A&A...465..965A}) are  displayed by filled circles, among which V$_{11}$ (labelled with `11'). Dashed lines show the theoretical $\gamma$\,Dor instability strip for $\alpha_{\rm MLT}$=2 as calculated by \cite{Dupt05}.}
\label{Fig1}
\end{figure}

The old open cluster NGC 2506 ($\alpha$, d)$_{2000}$=(08${\rm h}$
00$^{\rm m}$ 01$^{\rm s}$, -10$^{\circ}$ 46$^{\prime}$ 12$^{\prime \prime}$) has been reported to hold around 20 oscillating stars of
  different types, 15 of them being $\gamma$\,Dor candidates. The
  physical properties of the cluster are: distance = 3460 pc, age $\geq$ 2 Gyrand a metallicity of [Fe/H] = -0.20 $\pm$ 0.1
  dex (see \cite{2007A&A...465..965A} and references therin).

In Fig.~\ref{Fig1} we show these $\gamma$\,Dor candidates of NGC 2506
(filled circles) together with known bonafide (crosses) $\gamma$\,Dor stars
in the HR diagram. We also show the theoretical instability strip
obtained by \cite{Dupt05}.

In this study we concentrate on NGC 2506 V$_{11}$ (V = 15.454) which is labeled `11' in Fig.~\ref{Fig1}. Its global parameters are given by ${\rm T}_{\rm eff}$ = 7080 $\pm$ 150 K and Log (L/L$_{\sun}$) = 0.81 $\pm$ 0.03. It has been chosen as a first target because it shows the highest frequencies: 1.165, 1.270 and 1.400 (d$^{-1}$), respectively.

\section{Modelling of NGC 2506 V$_{11}$ using global parameters}

\begin{table}
\centering
\caption{Minimum and maximum values of the physical parameters of NGC 2506 V$_{11}$ matching its
observational parameters, with ages larger than 1800 Myr}
\label{tab1}
\begin{tabular}{cccccccc}\hline
M  & ${\rm T}_{\rm eff}$  &  [Fe/H]  &  L  &  Age & $\alpha_{\rm MLT}$ & $\alpha_{\rm ov}$ \\
(${\rm M}_{\sun}$) & (K) & (dex) & (${\rm L}_{\sun}$) & (Gyr) & & & \\
\hline \hline
1.25 & 6930 & -0.52 & 6.03 & 1.80 & 0.5 & 0.1\\
1.45 & 7229 & -0.12 & 6.92 & 3.12 & 1.5 & 0.3\\
\hline
\end{tabular}
\end{table}

The first step in the modelling of V$_{11}$ was the selection of
  those models matching the global parameters of the
  star. Obviously, stellar models could differ considering the
specific values chosen for these global parameters (${\rm T}_{\rm eff}$,
Metallicity, Log g, etc.) or for the values of the stellar
modelling such as the Mixing-Length parameter $\alpha_{\rm MLT}$ or
overshooting $\alpha_{\rm ov}$, etc.

We have constructed a   grid of models, using the CESAM equilibrium code (\citealt{Morel08}),
  varying the mass (in the range [1.25, 2.10] $\rm{M}_\odot$ with
  steps of $0.01 \rm{M}_\odot$), the metallicity (with values
  [M/H]=0.08, $-0.12$, $-0.32$and $-0.52$), the mixing-Length
  parameter (values $\alpha_{\rm MLT}$= 0.5, 1and 1.5) and 
  overshooting (values $\alpha_{\rm ov}$= 0.1, 0.2and 0.3). Standard physical inputs
  for $\gamma$\,Dor and $\delta$\,Scuti stars are used (see \citealt{Casas06, Casas09} for details). 
The pulsational observables have been calculated using the GraCo pulsational
code (\citealt{Moya04}; \citealt{Moya08a}). Both codes (CESAM and GraCo) have been
part of the ESTA group for comparison of codes (\citealt{Lebreton08, Moya08b}).

The number of models lying in the error box and fulfilling the cluster membership
  (sub-solar metallicity and age $>$ 1800 Myr), is huge (1210 in our
  grid). The minimum and maximum values obtained are displayed in
  Table~\ref{tab1}. The sub-solar metallicity constraint is automatically
  fulfilled when the age restriction is imposed, that is, there are no
  models of solar metallicity with an age larger that 1800 Myr and
  the global parameters of NGC 2506 V$_{11}$. This agrees with the global
  determination for the cluster.

\section{Theoretical Results - FRM}

\begin{table*}
\centering
\caption{Best models fitting the observed frequencies, with the mode identification and $I_{\rm{obs}}$ predicted by the FRM.}
\label{tab2}
\begin{tabular}{ccccccccccc}\hline
 & Model &   &         & n$_{1}$ & $\ell_{1}$ & n$_{2}$ & $\ell_{2}$ & n$_{3}$  & $\ell_{3}$ & $\chi^{2}$ \\
M/M$_{\sun}$ & [Fe/H] & $\alpha_{\rm MLT}$ & $\alpha_{\rm ov}$  &         &            &         &            &          &   & \\
\hline\hline
1.38 & -0.32 & 1.5 & 0.1 & -50 & 2 & -46 & 2 & -24 & 1 & 1.598E-06\\
1.36 & -0.32 & 1.5 & 0.1 & -49 & 2 & -45 & 2 & -41 & 2 & 2.563E-06 \\
1.27 & -0.52 & 1.5 & 0.2 & -30 & 1 & -48 & 2 & -25 & 1 & 6.073E-06 \\
1.36 & -0.32 & 1.5 & 0.2 & -48 & 2 & -44 & 2 & -23 & 1 & 6.861E-06 \\
1.28 & -0.52 & 1.5 & 0.1 & -56 & 2 & -30 & 1 & -47 & 2 & 8.374E-06 \\
1.33 & -0.32 & 1.5 & 0.2 & -48 & 2 & -44 & 2 & -40 & 2 & 9.794E-06 \\
\hline
\end{tabular}
\end{table*}

 The next step in our study was the inclusion of the asteroseismic
  constraints. To do so, we followed a self-consistent procedure
  presented by Moya et al. (2008) used in different asteroseismic
  studies of $\gamma$\,Doradus stars (\citealt{2006A&A...450..715R,2006A&A...456..261R,2008A&A...489.1213U}) and recently applied to the
  determination of the physical characteristics of the first planetary
  system observed by direct imaging, HR\,8799 (\citealt{hr8799a, hr8799b}).

  We first applied the FRM. This method provides sets of possible
mode identifications (n, $\ell$) for the three observed frequencies
and the corresponding value for the Brunt-V\"{a}is\"{a}l\"{a} integral
(I$_{obs}$). Adding this new constraint, the number of accepted models
is reduced by 70 $\%$. Then, we searched for those models
matching the observed frequencies with the predicted mode identification
and $I_{\rm{obs}}$. The model selection was done using $\chi^{2}$
statistics. The total number of modes selected is 50, for which the
standard deviation of the frequency fit is lower than $0.5\,\mu{\rm Hz}$.
Table~\ref{tab2} shows the models with the best $\chi^{2}$ and the
mode identification predicted.

\section{Theoretical Results - TDC}

\begin{figure}[t]
\centering
\includegraphics[width=82mm,height=75mm]{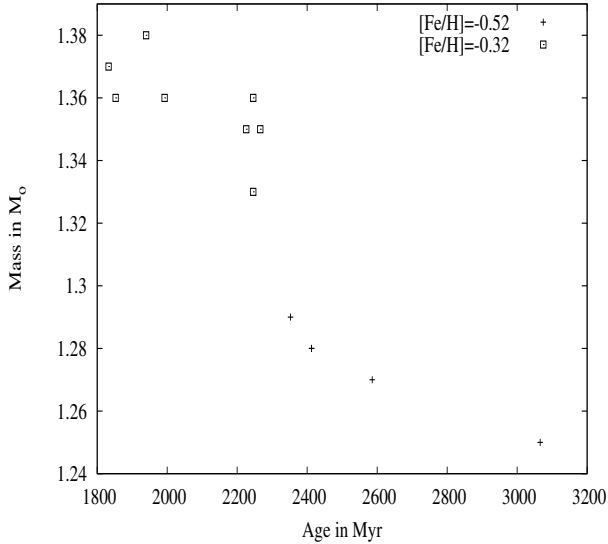}
\caption{Age - Mass diagram of the models fulfilling all the
  observational constraints. The metallicity of each model is also
  shown.}
\label{label2}
\end{figure}

The second step of our procedure is the study of the energy
  balance of the theoretical modes using TDC. We have accepted
those models selected in the previous step that predict over-stable
observed frequencies. The models fulfilling all the asteroseismic
constraints are drastically reduced to those displayed in
Fig.~\ref{label2}. In this figure we show the Age - Mass values
  of these models together with their metallicity. The minimum and
  maximum values obtained are displayed in Table~\ref{tab3}.

A first consequence of this study is that models with an age around 1800 -
  2000 Myr have an internal metallicity similar to that observed for
  the cluster ([Fe/H]=-0.32), masses in the range [1.33,1.38]
  M$_{\odot}$ and radius around 1.77 R$_{\odot}$. Lower metallicity
  models can also fulfill all the observational constraints but with
  ages larger than expected for the stars of NGC 2506.

\section{Conclusions}
With this prospective study we showed that the application of FRM and TDC to the cluster member $\gamma$\,Dor candidate NGC 2506 V$_{11}$ enhances significantly our determination of the mass, radius and age. Specifically, its
metallicity is confirmed to be sub-solar. The energy transport efficiency of the outer convective zone is high ($\alpha_{\rm MLT}=1.5$). These results must be confirmed with the extension of this study to other $\gamma$\,Dor stars of this cluster.

This is a first step of a more ambitious project. In the near future, we plan to expand the study to other
variable members of the NGC 2506 cluster. A better knowledge of variable membership (\citealt{Ottosen08}) and their global parameters will improve our knowledge on $\gamma$\,Dor pulsators and probe the self-consistency of our procedures.

\begin{table*}
\centering{
\caption{Minimum and maximum values of the physical parameters of NGC 2506 V$_{11}$ fulfilling all the constraints}
\label{tab3}
\begin{tabular}{lccccccc}\hline
Mass  & ${\rm T}_{\rm eff}$(K)  &   [Fe/H]  &  L/L$_{\sun}$  &  R/R$_{\sun}$   &  Age (Myr) & $\alpha_{\rm MLT}$ & $\alpha_{\rm ov}$ \\
\hline \hline
1.25 -- 1.29 & 6997 & -0.52 & 6.07 & 1.6   & 1832 & 1.5 & 0.1\\
1.33 -- 1.38 & 7223 & -0.32 & 6.83 & 1.774 & 2585 & 1.5&  0.3\\
\hline
\end{tabular}}
\end{table*}

\acknowledgements This work was supported by the European Helio- and Asteroseismology Network (HELAS), a major international collaboration funded by the European Commission's Sixth Framework Programme. AG is supported by grant with reference number SFRH/BPD/41270/2007 and project PTDC/CTE-AST/098754/2008 from FCT/MCTES, Portugal. AM acknowledges financial support from a Juan de la
Cierva contract of the Spanish Ministry of Science and Innovation. This research has been in part funded by Spanish grants
ESP2007-65475-57-C02-02, CSD2006-00070, CAM/PRICIT-S2009ESP-1496 and
ESP2007-65480-C02-01. JCS acknowledges support from the ``Instituto de Astrof\'{\i}sica de Andaluc\'{\i}a
(CSIC)'' by an ``Excellence Project post-doctoral fellowship'' financed by the Spanish ``Conjerer\'{\i}a
de Innovaci\'on, Ciencia y Empresa de la Junta de Andaluc\'{\i}a'' under project ``FQM4156-2008''.


\begin{thebibliography}{}

\bibitem[Arentoft et al.(2007)]{2007A&A...465..965A} 
Arentoft, T., De Ridder, J., Grundahl, F. et al.: 2007, A\&A 465, 965 

\bibitem[Casas et al.(2006)]{Casas06} 
Casas, R. et al.: 2006, A\&A 455, 1019

\bibitem[Casas et al.(2009)]{Casas09} 
Casas, R. et al.: 2009, \apj697, 522

\bibitem[Creevey et al.(2009)]{2009A&A...507..901C}
Creevey, O.L., Uytterhoeven, K., Mart{\'i}n-Ruiz, S., et al.: 2009, A\&A 507, 901 

\bibitem[Dupret et al.(2005)] {Dupt05} 
Dupret, M.-A., Grigahc{\`e}ne, A., Garrido, R., Gabriel, M., Scuflaire, R.: 2005, A\&A 435, 927

\bibitem[Grigahc{\`e}ne et al.(2005)]{2005A&A...434.1055G} 
Grigahc{\`e}ne, A., Dupret, M.-A., Gabriel, M., Garrido, R., Scuflaire, R.: 2005, A\&A 434, 1055 

\bibitem[Guzik et al.(2000)]{Guzik00} 
   Guzik, J.A., Kaye, A.B., Bradley, P.A., Cox, A.N., Neuforge, C.: 2000, \apj542, 57

\bibitem[Henry et al.(2007)]{Hen07} 
Henry, G.W., Fekel, F.C., Henry, S.M.: 2007, \aj133, 1421

\bibitem[Lebreton et al.(2008)]{Lebreton08} 
Lebreton, Y., et al. 2008, Ap\&SS 316, 1

\bibitem[Mart{\'{\i}}n \& Rodr{\'{\i}}guez(2002)]{2002ASPC..259..152M} 
Mart{\'{\i}}n, S., \& Rodr{\'{\i}}guez, E.: 2002, IAU Colloq.~185: Radial and Nonradial Pulsationsn as Probes of Stellar Physics, 259, 152 

\bibitem[Morel \& Lebreton(2008)]{Morel08} 
Morel P., Lebreton Y.: 2008, Ap\&SS 316, 61 

\bibitem[Moya, Garrido \& Dupret(2004)]{Moya04} 
Moya, A., Garrido, R. Dupret, M.-A.: 2004, A\&A 414, 1081

\bibitem[Moya et al.(2005)] {Moya05} 
Moya, A., Su{\'a}rez, J.C., Amado, P.J., Mart{\'i}n-Ru{\'i}z, S., Garrido, R.: 2005, A\&A 432, 189

\bibitem[Moya et al.(2008)]{completo} 
Moya A., et al.: 2008, \an329, 541

\bibitem[Moya \& Garrido(2008)]{Moya08a} 
Moya, A., Garrido, R.: 2008, Ap\&SS 316, 129

\bibitem[Moya et al.(2008)]{Moya08b} 
Moya, A., et al.: 2008, Ap\&SS 316, 231

\bibitem[Moya et al.(2010a)]{hr8799a} 
Moya A., et al.: 2010a, submitted to \mnras

\bibitem[Moya et al.(2010b)]{hr8799b} 
Moya A., et al.: 2010b, submitted to MNRAS Letters

\bibitem[Ottosen(2008)]{Ottosen08}
Ottosen, T.A.: 2008, Studies of the open cluster NGC 2506 based on VLT observations, Bachelor project, Aarhus University

\bibitem[Rodr{\'{\i}}guez et al.(2006a)]{2006A&A...450..715R} 
Rodr{\'{\i}}guez, E., et al.: 2006, A\&A 450, 715 

\bibitem[Rodr{\'{\i}}guez et al.(2006b)]{2006A&A...456..261R} 
Rodr{\'{\i}}guez, E., et al.: 2006, A\&A 456, 261 

\bibitem[Su{\'a}rez et al.(2002)]{2002A&A...390..523S} 
Su{\'a}rez, J.C., Michel, E., P{\'e}rez Hern{\'a}ndez, F., Lebreton, Y., Li, Z.~P., Fox Machado, L.: 2002, A\&A 390, 523 

\bibitem[Su{\'a}rez et al.(2005)]{Suarez05} 
Su{\'a}rez, J.C., Moya, A., Mart{\'i}n-Ru{\'i}z, S., Amado, P.~J., Grigahc{\`e}ne, A., Garrido, R. 2005: A\&A 443, 271

\bibitem[Su{\'a}rez et al.(2007)]{2007MNRAS.379..201S} 
Su{\'a}rez, J.C., Michel, E., Houdek, G., P{\'e}rez Hern{\'a}ndez, F., Lebreton, Y.: 2007, \mnras379, 201 

\bibitem[Smeyers \& Moya(2007)] {Smey07} 
Smeyers, P., Moya, A.: 2007, A\&A 465, 509

\bibitem[Uytterhoeven et al.(2008)]{2008A&A...489.1213U} 
Uytterhoeven, K., et al.: 2008, A\&A 489, 1213 


\end{thebibliography}
\end{document}